\documentclass{article}

\usepackage{PRIMEarxiv}

\usepackage[utf8]{inputenc} 
\usepackage[T1]{fontenc}    
\usepackage{hyperref}       
\usepackage{url}            
\usepackage{booktabs}       
\usepackage{amsfonts}       
\usepackage{amsmath}
\usepackage{nicefrac}       
\usepackage{microtype}      
\usepackage{lipsum}
\usepackage{graphicx}
\graphicspath{{media/}}     

\usepackage{graphicx}
\usepackage{subcaption}
  
\title{A Learned Born Series for Highly-Scattering Media
}

\author{
  Antonio Stanziola, Simon Arridge, Ben T. Cox, Bradley E. Treeby \\
  University College London\\
  Gower Street, London\\
  WC1E 6BT, UK \\
  \texttt{\{a.stanziola, s.arridge, b.cox, b.treeby\}@ucl.ac.uk} \\
}

\begin{document}
\maketitle

\begin{abstract}
A new method for solving the wave equation is presented, called the learned Born series (LBS), which is derived from a convergent Born Series but its components are found through training. The LBS is shown to be significantly more accurate than the convergent Born series for the same number of iterations, in the presence of high contrast scatterers, while maintaining a comparable computational complexity. The LBS is able to generate a reasonable prediction of the global pressure field with a small number of iterations, and the errors decrease with the number of learned iterations.
\end{abstract}


\section{Introduction}
Computational solvers for the Helmholtz equation with heterogeneous material properties are used in many fields that require single frequency wave simulations, including optics, seismology, acoustics, and electromagnetics. One approach is to compute the terms of the Born series \cite{born1926quantenmechanik} through the recursive application of the known Green's operator for a homogeneous medium. However, in its conventional form the Born series converges only for low contrast scattering potentials - a major limitation in many practical applications. As an example, the application that has motivated this work is the propagation of ultrasound waves through the skull, which has a sound speed about double that of the surrounding soft tissue. A convergent Born series (CBS) was proposed by Kleinman et al.\ based on a generalised overrelaxation method \cite{kleinman1988iterative,kleinman1990convergent}. More recently, Osnabrugge et al.\  introduced another CBS through the use of a preconditioner based on the scattering potential \cite{OSNABRUGGE2016113}, which has since found many applications 
\cite{lee2022inverse,jakobsen2020homotopy,kruger2017solution,kaushik2020convergent}. However, for strongly scattering media many iterations are needed, as the convergence rate is limited by the range of spatial wavenumbers in the problem. To overcome the slow convergence rate, we present a learned Born series (LBS), in which a generalised preconditioner and the modified Green's operator are found, for a given number of unrolled iterations \cite{monga2021algorithm}, through training. We show that the convergence rate in the presence of high contrast scatterers is considerably improved.


\section{Learned Born Series}


We aim to solve the Helmholtz differential equation $(\nabla ^2 + k^2)u = -s$, where $k = \omega / c_0$ is the heterogeneous wavenumber, $\omega$ is the angular frequency, and $c_0$ is the spatially-varying sound speed. To solve this equation, following Osnabrugge et al. \cite{OSNABRUGGE2016113}, we start by defining the scattering potential $v = k^2 - \kappa^2 - i\varepsilon$, where $\kappa^2,\varepsilon \in \mathbb{R}_+$ are positive scalars, and rewrite the differential equation as
\begin{equation}
\mathcal Lu = (\nabla^2 +\kappa^2 + i\varepsilon)u = -s - vu.
\label{eq:modifiedhelmholtz}
\end{equation}
We denote with $\mathcal G$ the Green's operator associated with the operator $\mathcal L$, that is
\begin{equation}
    \mathcal G = \mathcal{F}^{-1} (p^2 - \kappa^2 - i \varepsilon)^{-1} \mathcal F,
\end{equation}
where $\mathcal F$ is the Fourier transform operator and $p^2 = p_x^2 + p_y^2$ is the squared magnitude of the Fourier transformed coordinates. Applying the Green's operator to both sides of Eq.\ \eqref{eq:modifiedhelmholtz} leads to $u = \mathcal Gs + \mathcal G v u$. Repeated substitution of the right-hand-side into $u$ leads to the classical form of the Born series:
\begin{equation}
    u = \left[1 + \mathcal G v + (\mathcal G v)^2 + (\mathcal G v)^2 + \dots \right]\mathcal Gs = \sum_{n=0}^\infty (\mathcal G v)^n \mathcal G s,
\end{equation}
under the assumption that the series converges. As mentioned in Sec.\ 1, this assumption is satisfied only for low scattering media. Osnabrugge et al.\ \cite{OSNABRUGGE2016113} proposed the preconditioner $q = iv/\varepsilon$, leading to the CBS
\begin{equation}
    u = \left[1 + \mathcal M + \mathcal M^2 + \mathcal M^3 + \dots \right]q\mathcal Gs, \qquad \mathcal M = q \mathcal G v + (1-q),
    \label{eq:ConvBornSeries}
\end{equation}
which is convergent for scattering media with arbitrarily-strong sound speed contrasts for an appropriately chosen $\varepsilon$. Despite the convergence guarantees of Eq.\ \eqref{eq:ConvBornSeries}, the convergence rate is inversely proportional to the maximum wavenumber difference between any two points in the domain. This means that for media with sound-speed high contrast, a large number of iterations are necessary.

With the aim of reducing the number of iterations required to solve problems in highly-scattering media, we modify the CBS as follows. First, note that the series in Eq.\ \eqref{eq:ConvBornSeries} can be rewritten as
\begin{equation}
    u_n = u_{n-1} + w_n, \qquad w_{n} = \mathcal M w_{n-1},
\end{equation}
with $u_0 = w_{0} = q\mathcal G s$. 
To develop a learned version of Eq.\ \eqref{eq:ConvBornSeries}, we transform the linear operator $\mathcal M$ into a non-linear one in the form 
\begin{equation}
\mathcal M = q \mathcal G v + (1-q) \quad \rightarrow \quad    \hat{\mathcal M}_{\theta} = (A_{\theta} \hat{\mathcal G}_{\theta} B_{\theta} + C_{\theta}).
    \label{eq:LearnedBorn1}
\end{equation}
Here, $A$ is the learned version of the preconditioner $q$, $B$ is the learned version of the scattering potential $v$, and $C$ is now a general additive term (that is not enforced to be $1 - A$). Each of $A, B, C$ are the outputs of a neural network dependent on the input sound speed and the spatial location maps, i.e.,
\begin{equation}
A_{\theta} = f^{A}(c, \mathbf x; \theta), \quad 
B_{\theta} = f^{B}(c, \mathbf x; \theta), \quad 
C_{\theta} = f^{C}(c, \mathbf x; \theta),
\end{equation}
where, to simplify notation, we use $\theta$ to denote the learnable parameters for each operator at each iteration, without implying that they are the same set.
The learned fields $(A,B,C)$ are returned by three fully connected neural networks (FCNN) acting channelwise on the $(c, \mathbf x)$ map, also known as 1$\times$1 convolutions, with two hidden layers. The size of the hidden layers is equal to the number of channels.
Similar to convolutional networks, we generalise Eq.\ (\ref{eq:LearnedBorn1}) to $c$ channels by associating the action of each $(A,B,C)$ field to a matrix product; for example if $\mathbf u \in \mathbb R^c = (u_1, u_2, \dots, u_c)$, then $(A\mathbf u)_i = \sum_j A_{ij} u_j$. Note that the values of $(A,B,C)$ change for each pixel location. 

\begin{figure}[t]
    \centering
    \begin{minipage}{.55\linewidth}
        \begin{subfigure}[t]{.9\linewidth}
            \includegraphics[width=\textwidth]{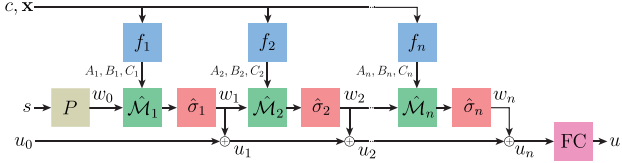}
            \caption{\label{fig:diagram} Block diagram for the learned Born series (LBS) with $n$ iterations.}
        \end{subfigure} \\
        \begin{subfigure}[t]{.9\linewidth}
            \includegraphics[width=\textwidth]{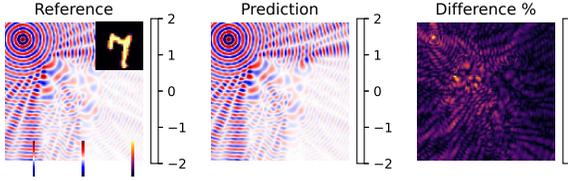}
            \caption{
            \label{fig:example_against_reference} Example result against reference.}
        \end{subfigure} \\
        \begin{center}
            \begin{subfigure}[t]{.6\linewidth}
                \includegraphics[width=\textwidth]{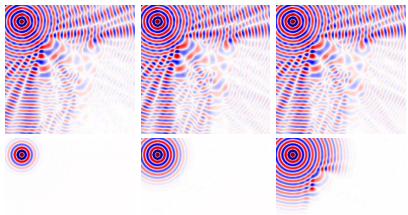}
                \caption{\label{fig:iterations_images} Solution for 6, 12 and 24 iterations. The LBS is shown on top, while the CBS is shown at the bottom.}
            \end{subfigure}
        \end{center}
    \end{minipage}
    \begin{minipage}{.42\linewidth}
        \begin{subfigure}[b]{.9\linewidth}
            \includegraphics[width=\textwidth]{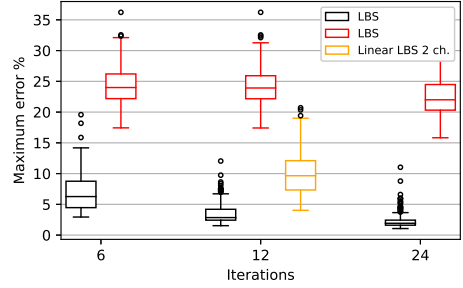}
            \caption{Maximum error over the test set for different number of iterations.}
            \label{fig:boxplots}
        \end{subfigure} 
            \begin{subfigure}[t]{.9\linewidth}
                \hspace*{-6mm} 
                \includegraphics[width=1.2\textwidth]{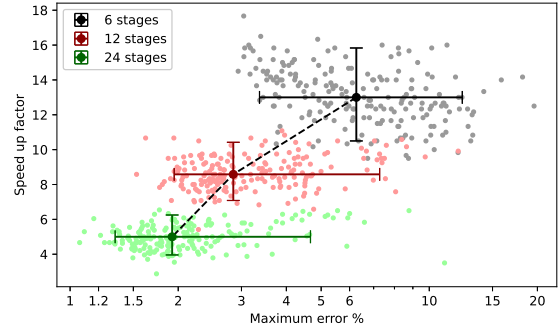}
                \caption{Number of iterations required by CBS over number of LBS stages, for a similar error, for the different number of stages.
                \label{fig:speed_up}}
            \end{subfigure}
        \end{minipage}
    \caption{Results on the test set.}
    \label{fig:test_results}
\end{figure}

We also make the wavenumber shift of the Green's function, $\kappa^2 + i\varepsilon$, learnable. To avoid dividing by zero during training, instead of directly learning $\kappa^2$ and $\varepsilon$, we  parameterise the Green's operator as
\begin{equation}
    \hat{\mathcal G}_{\theta} = \mathcal F^{-1} (\alpha p^2 - \kappa^2 - i)^{-1}  \mathcal F,\quad \theta = \{\alpha,\kappa^2\}
\end{equation}
where $\alpha$ and $\kappa^2$ are learnable parameters, which is equivalent up to a scaling factor (that is absorbed by the $A$ term). 
Similarly, the input to the Green's operator is an $n-$dimensional field, therefore the Green's function is a mapping $\mathbb R^c \to \mathbb R^c$, thus $\alpha, \kappa^2$ are $c\times c$, where $c$ is the number of channels. Lastly, we apply a projected non-linearity of the form $\hat \sigma(u) = D_1\sigma(D_2 u)$, where $D_1,D_2$ are two dense layers acting channel-wise and $\sigma$ is a non-linear function, in our case a GeLU function \cite{hendrycks2016gaussian}. The final learned iteration is therefore
\begin{equation}
    u_n = u_{n-1} + w_n, \qquad w_{n} = \hat \sigma_n \left(\hat{\mathcal M}_n w_{n-1}\right), \qquad n = 1 \hdots m
    \label{eq:single_iteration}
\end{equation}
where we have used the subscript $n$ to highlight that $\hat \sigma$ and $\hat{\mathcal M}$ have different parameters at each iteration. We set $u_0 = 0$.
We note the similarity between Eq.\ \eqref{eq:single_iteration} to the Fourier Neural Operators (FNO) framework \cite{li2020fourier}, where the network is set up as the sum of a Fourier filter and a local convolution operator as
\begin{equation}
    u_{n+1} = \sigma\left(\mathcal N u_n\right), \quad \mathcal N =  \mathcal F^{-1} K \mathcal{F} + C
    \label{eq:FNO1}
\end{equation}
where $K$ is a learnable Fourier filter and $C$ is a learnable local convolution filter. 
Similarly to the FNO, we pad the input by a fixed amount to accommodate the artifacts generated by the periodicity assumed by the discrete Fourier transform, and afterwards unpad the output. A block diagram of the network is given in Fig.\ \ref{fig:diagram}.

\section{Training and Results}
A dataset was constructed from 2000 MNIST \cite{deng2012mnist} samples upsampled to 128$\times$128 pixels, which were rescaled and used as sound speed distributions. The simulations were performed in normalised units with a background sound speed of 1 m.s$^{-1}$, maximum sound speed of 2 m.s$^{-1}$, and $\omega = 1$ rad.s$^{-1}$ (giving $2\pi$ grid points per wavelength). The dataset was split between train, validation and test in 80\%, 10\% and 10\% -sized subsets. 
For each sound speed map, a reference solution was generated using the Helmholtz solver in j-Wave \cite{stanziola2022jwave} using GMRES with a fixed source position. The loss function was the mean squared error against the reference solution.
All networks were trained with a batch-size of 16 and learning rate of $3\cdot 10^{-3}$: the network corresponding to the epoch with the lowest validation loss was saved, although we didn't observe any over-fitting during training.
The network was trained for 6, 12 and 24 unrolled iterations, with $c=8$. We also trained an extremely light version with $c=2$ (matching the number of components on a complex field, as in the original CBS) with linear activation functions.


The simulated pressure field for a representative example from the test set using a network trained with 12 iterations is shown in Fig.\ \ref{fig:example_against_reference}. Even with a very small number of iterations, the LBS is able to generate a reasonable prediction of the global pressure field. The errors against j-Wave for the whole test set are shown in Fig.\ \ref{fig:boxplots}. As expected, the errors decrease with the number of learned iterations. Compared to the CBS run for the same number of iterations, the error is significantly reduced. This is because the pseudo propagation speed in the CBS is limited due to the strong contrast, as shown in Fig.\ \ref{fig:iterations_images}. 
The outliers correspond to speed of sound maps that exhibit resonant modes, 
eg. ``0'' digits for which the field exhibited a mode-like structure 
similar to whispering gallery modes \cite{stanziola2021helmholtz}. The error for the CBS was also proportionally higher for these examples. To match the same error on the test set of the LBS using 6, 12, and 24 iterations, on average the CBS required 13.1, 8.6, and 5.0 times more iterations, respectively (see Fig.\ \ref{fig:speed_up}). Even when the number of channels is reduced to 2 (equivalent to the complexity of the CBS), the convergence of the LBS is still much faster than the CBS (see yellow boxplot in Fig.\ \ref{fig:boxplots}).

\section{Summary and Discussion}

We present a learned Born series, inspired by the convergent Born series presented in \cite{OSNABRUGGE2016113}, that significantly improves convergence rates for single-frequency wave simulations with strong sound speed contrast. While for channel counts $c > 2$ the LBS requires more FFTs at each layer compared to the CBS, such FFTs can be computed in a parallel fashion, which still makes the method more efficient in practice.

In the future, it would be interesting to explore several potential extensions of this work. One possibility is to train on sub-sampled problems that are otherwise unsolvable with classical methods, in order to further reduce the computational complexity of the LBS. This approach could have applications in areas such as transcranial treatment planning and full waveform inversion. Another interesting direction would be to apply the method to other PDEs for which a scattering potential can be defined, such as the Schrödinger equation. 

One potential way to improve the LBS would be to substitute the fixed padding with a learnable ``PML-like'' layer, instead of feeding the spatial grid $\mathbf x$. This could lead to an architecture that is resolution and grid-size independent, making it a neural operator as described in \cite{kovachki2021neural}. It would also be interesting to train the method as an iterative solver, as described in \cite{stanziola2021helmholtz}.
Lastly, it would be interesting to explore whether the LBS can be extended to heterogeneous density. While the method presented in this paper does not support it, the CBS does not either. Allowing so would open up new applications in areas where the density of the medium being modeled is non uniform, such as ultrasound applications with the presence of bone or geophysics.

\section*{Acknowledgments}

This work was supported by the Engineering and Physical Sciences Research Council (EPSRC), UK, grant numbers EP/S026371/1, EP/T022280/1, and EP/W029324/1.

\bibliographystyle{unsrt}  
\bibliography{references}

\end{document}